# Design of a 325MHz Half Wave Resonator prototype at IHEP

ZHANG Xinying(张新颖)[1,2] PAN Weimin(潘卫民)[2] WANG Guangwei(王光伟)[2] XU Bo(徐波)[2] ZHAO Guangyuan(赵光远)[2] HE Feisi(贺斐思)[2] LI Zhongquan(李中泉)[2] MA Qiang(马强)[2] DAI Jin(戴劲)[2] CHEN Xu(陈旭)[2] MI Zhenghui(米正辉)[2] SHA Peng(沙鹏)[2] LIN Haiying(林海英)[2] WANG Qunyao(王群要)[2] LIU Yaping(刘亚萍)[2] XUE Zhou(薛舟)[1,2] HUANG Xuefang(黄雪芳)[1,2] SUN Yi(孙毅)[2]

1 (University of the Chinese Academy of Sciences, Beijing 100049, China)
2 (Institute of High Energy Physics, CAS, Beijing 100049, China)

**Abstract**：A 325MHz $\beta=0.14$ superconducting half wave resonator(HWR) prototype has been developed at the Institute of High Energy Physics(IHEP), which can be applied in continuous wave (CW) high beam proton accelerators. In this paper, the electromagnetic (EM) design, multipacting simulation, mechanical optimization, and fabrication are introduced in details. In vertical test at 4.2K, the cavity reached $E_{acc}$=7MV/m with $Q_0$=1.4×10$^9$ and $E_{acc}$=15.9MV/m with $Q_0$=4.3×10$^8$.

## 1. Introduction

Superconducting(SC) HWR is an accelerating structure used for low and medium $\beta$ beam energy. Compared with SC quarter wave resonator (QWR), the HWR has more symmetry structure, which conceals the vertical beam steering effect and obtains about 2 times larger power at the same accelerating voltage. Compared to spoke type cavity with similar $\beta$, HWR can be more compact and higher gradient and better mechanical stability. Since these advantages, more and more new facilities propose using HWRs to accelerate the proton beam. The Project X at Fermi National Accelerator Laboratory proposed using HWRs with 162.5MHz [1]. The driver of the International Fusion Material Irradiation Facility (IFMIF) at CEA-Saclay also proposed using HWRs with 175MHz [2]. The driver accelerator for the Facility for Rare Isotope Beams (FRIB) will use $\beta$=0.29 HWRs and $\beta$=0.53 HWRs with 322MHz [3]. A high beam proton accelerator for Accelerator Driven Sub-critical System (C-ADS) plans to use HWRs with 162.5MHz [4]. A 325MHz HWR prototype has been developed at IHEP. The main parameters are summarized in table 1.

Table1: Main parameters of 325MHz HWR

| Requirements | Description |
|---|---|
| Particle type | Proton |
| Frequency | 325MHz |
| $\beta$ | 0.14 |
| Operating mode | CW |
| $R_{aperture}$ | 35mm |
| Beam current | 10mA |

## 2. Design

The EM design, multipacting, and mechanical design was optimized, meanwhile special attention was paid to make the design compatible with established techniques for cavity fabrication and surface preparation, maximizing the probability of reliable performance.

### 2.1 EM design

In EM design, $E_{peak}/E_{acc}$, $B_{peak}/E_{acc}$, R/Q, and G should be highly concerned. The goal of EM design is to optimize the cavity geometry to minimize the value of $E_{peak}/E_{acc}$ to avoid the field emission, and to minimize the value of $B_{peak}/E_{acc}$ to maintain

zhxy@ihep.ac.cn

superconductivity.

For the cavity's mechanical stability, the spherical shape of the cavity was chosen, and all the interfaces with ports are circular, which makes all electron beam welding (EBW) easier. The cover of the cavity was dome shape, which makes cavity more rigid and minimizes multipacting effect. The optimized cavity geometry is shown in Fig.1 and cavity parameters are summarized in Table 2.

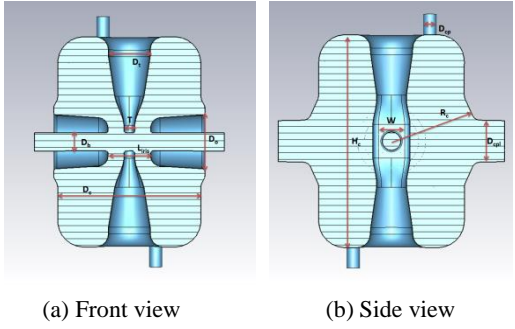

(a) Front view    (b) Side view

Fig. 1: The section view of the HWR.

Table2: The optimized geometrical parameters

| Geometrical parameters | Value(mm) |
|---|---|
| Cavity height $H_c$ | 406 |
| Cavity diameter $D_c$ | 274 |
| Cavity center radius $R_c$ | 160 |
| Inner top diameter $D_t$ | 80 |
| Racetrack thickness T | 20 |
| Racetrack width W | 50 |
| Iris length Liris | 82 |
| Beam prot inner diameter $D_b$ | 35 |
| Beam prot outer diameter $D_o$ | 117 |
| Coupler prot diameter $D_{cpl}$ | 80 |
| Cleaning port diameter $D_{cl}$ | 25 |

The specifications of EM design obtained with the simulations by 3D solver CST-MWS are listed in Table3. The finalized electromagnetic field distribution is shown in Fig.2, and the finalized longitudinal voltage along the axis of the HWR is shown in Fig.3.

Table3: The optimized RF parameters of the HWR

| RF parameters | Result |
|---|---|
| $E_{peak}/E_{acc}$ | 4.2 |
| $B_{peak}/E_{acc}$ | 4.9 mT/(MV/m) |
| R/Q | 195 Ω |
| G | 74 Ω |

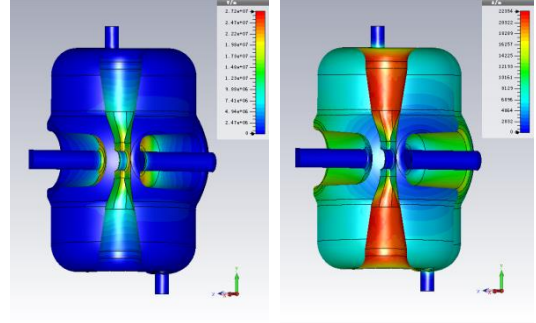

(a) E-field distribution    (b) H-field distribution

Fig. 2: Electromagnetic field distribution of the HWR.

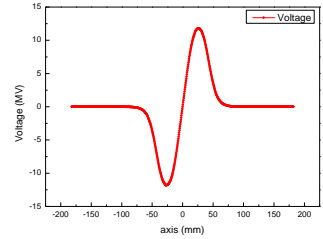

Fig. 3: Longitudinal voltage along the axis.

**2.2 Multipacting**

Multipacting in RF structures is a resonant process in which a large number of electrons build up a multipacting discharge, absorbing RF power so that it becomes impossible to increase the cavity fields by raising the incident power[2]. A proper cavity geometry and perfect surface preparation can stop multipacting from happening in the cavity. The multipacting analysis is done by using Track3P module developed by SLAC. Typical SEY (Secondary Emission Yield) curve for niobium of 300° bake out is shown in Fig.4, in which indicated the multipacting rang of kinetic energy is about from 70 eV to 1600 eV.

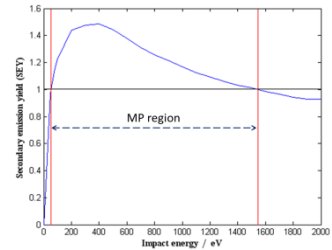

Fig. 4: SEY curve for Niobium of 300℃ bake out.

The multipacting of the 325MHz β=0.14

HWR prototype was simulated for $E_{acc}$ from 1MV/m to 11 MV/m. Resonant trajectories happen at two regions as shown in Fig.5, from 2.2 to 5.1 and from 8.7 to 9.4. There are no hard multipacting barrier happens at operating gradient 7MV/m.

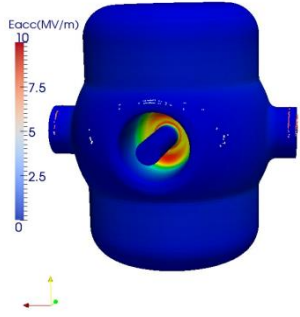

(a) The resonant points location

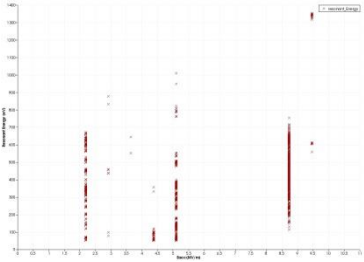

(b) The impact resonant energy VS. Eacc

Fig. 5: The multipacting results from Track3P.

**2.3 Mechanical design**

The mechanical performance of the 325MHz HWR prototype is optimized using SolidWorks CAD software and ANSYS code, which includes reasonable rigidity, the cavity's tuning range, acceptably low pressure sensitivity, Lorentz force detuning (LFD) verifying, and microphonic detuning. A stiffening ring was used to enforce the nose cup around beam ports. The thickness of the cavity wall is determined to be 3.2mm.

The stresses on the cavity was simulated and summarized in Table 4. The allowable stresses for niobium RRR300 based on the yield strength are 47MPa at room temperature (RT) and 212MPa at 4K. The results indicated that the cavity is safe by the evacuation, cool down and tuning condition. The Lorentz pressure in CW operation is negligible which is about 0.05MPa at $E_{acc}$=5MV.

Table4: The optimized stress of the HWR

| Parameter | Boundary | Stress (MPa) |
|---|---|---|
| Evacuation(1bar, RT) | ports fixed | 15.2 |
| | ports free | 20.1 |
| Cool down(4.2K) | beam ports fixed | 134 |
| Tuning(2kN, RT) | coupler ports free | 36 |

A constant pressure on the outside of the cavity wall will cause a displacement and frequency detuning. The resonant frequency (f) due to a change in ambient pressure (p) is dominantly linear, which is quantified by the coefficient $df/dp$. As beam pipe ports and coupler ports fixed, the $df/dp$ is -6.3Hz/mbar, and as beam pipe ports and coupler ports are free, the $df/dp$ is -95.3Hz/mbar. The ports fixed condition results reflects the real tuner influence more accurately. The deformation and stress distribution of the cavity at 1bar is shown in Fig. 6.

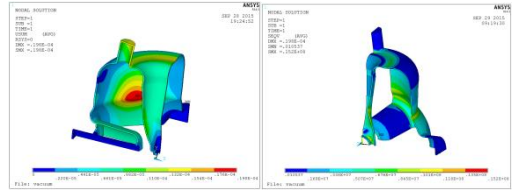

(a) Beam and coupler ports fixed

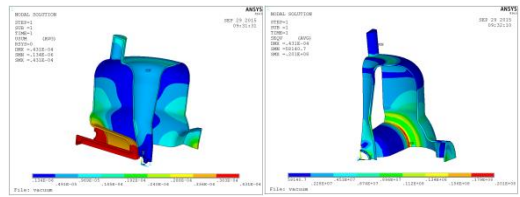

(b) Beam and coupler ports free

Fig. 6: The deformation and stress of the HWR under atmospheric loading.

The tuning force is applied on the flange of the beam pipe. The tuning range R and the tuning force F are related with the stiffness k and the tuning sensitivity $s$[5].

$$R = \frac{s}{k}F$$

The simulation results indicate $s =$

1.1MHz/mm, k = 17.3kN/mm. 1.6kN is needed for 100 kHz tuning range.

The interaction of the surface electromagnetic fields in a cavity with the induced surface currents and charges results in a Lorentz force on the cavity walls [6]. This pressure will results in a deformation of the cavity walls, and then causes the shift of the resonant frequency of the cavity. The LFD coefficient has been simulated and shown in Fig. 7. The $K_L = \Delta f / E_{acc}^2$ is -2.1 Hz/(MV/m)$^2$ with ports fixed and -12.7Hz/(MV/m)$^2$ with ports free.

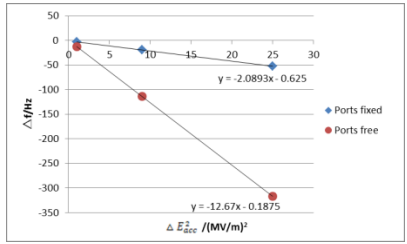

Fig. 7: The LFD coefficient of the HWR.

In order to get an accurate frequency (325MHz) in the 4.2 K, the change of frequency has been studied, and the results are summarized in Table 5. The buffer chemical polish (BCP) and cooling down to 4.2K will make the resonant frequency higher, while the evacuation and LFD will make the resonant frequency lower. So the aim frequency is 323.8MHz when the cavity has been fabricated. The Fig. 8 is the deformation of HWR when the temperature changes from 295K to 4.2K.

Table5: The frequency changes of the HWR

| Performance | Boundary | △f/kHz |
| --- | --- | --- |
| BCP(200 μ m) | ---- | +886 |
| Evacuation | ports fixed | -6.33 |
|  | ports free | -95.29 |
| $\varepsilon_{air} \rightarrow \varepsilon_{vacuum}$ | ---- | -94.65 |
| Cool down (to 4.2K) | ports fixed | +492.35 |
| LFD (operating $E_{acc}$) | ports fixed | -0.10 |
|  | ports free | -0.62 |

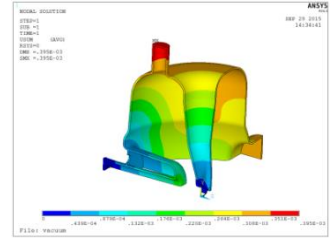

Fig. 8: The deformation of the HWR at cooling down.

The cavity was analyzed for the mechanical resonance modes. Low frequency modes around 250 Hz and below lead to microphonic resonances and must be avoided. Fig. 9 and Table 6 show the lowest mechanical frequency is 369Hz, indicates there is minimal danger to microphonic resonances.

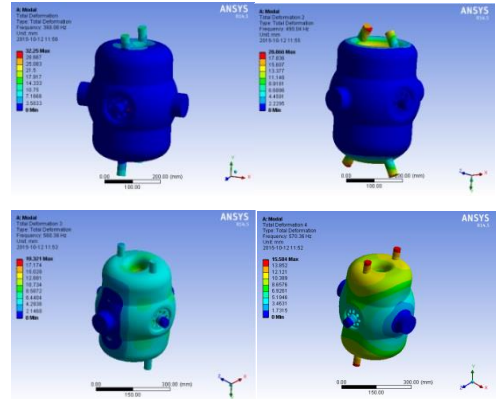

Fig. 9: The lowest eigenvector modal shapes.

Table6: The modal analysis

| Mode | Frequency(Hz) |
| --- | --- |
| 1 | 369 |
| 2 | 490 |
| 3 | 560 |
| 4 | 570 |

## 3. Fabrication

The cavity and stiffening rings are made of niobium RRR300, while the flanges are made of Nb-Ti alloy. An exploded view of 325MHz HWR is shown in Fig. 10.

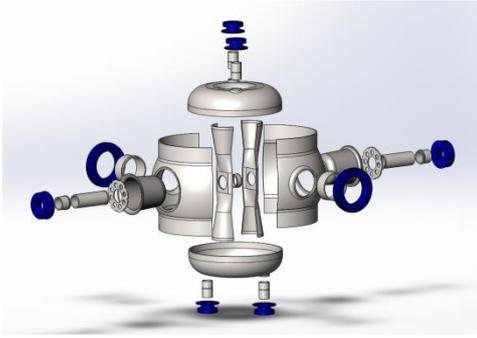

Fig. 10: The exploded view of the HWR.

The spinning forming, deep drawing, and bulging technology are used for the cavity's components. The outer conductors, covers, and nose cups are made integrally by spinning forming, which may reduce the number of welds. After spinning forming, the outer conductors are bulged to make the spherical curves of center part. The inner conductor and the holes in the covers are made by deep drawing technology. The pipes, stiffening rings, and flanges are made by machining with accuracy. The annealing of some components is necessary to eliminate the residual stress, which is caused by the mechanical machining, and it will be harmful to the machining dimension accuracy and assembling.

EBW is used to joint all of the HWR components together. The fabrication sequence is shown in Fig.11. Before final EBW, the inner surface of the cavity should be examined carefully, and confirmed the defects has been removed out. Every components were underwent a chemical polish to wipe off the oxide layer from the weld region. In order to guarantee the weld quality, the wall thickness of the weld region should be within 3.2 mm, and the thickness tolerance should be less than 0.1mm. The maximum docking circular seam gap is 0.1 mm. The estimated value of shrinkage due to the last EBW was 0.6mm. The bare HWR prototype is shown in Fig.12.

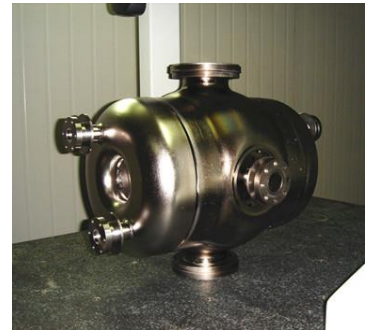

Fig. 12: The 325MHz bare HWR prototype.

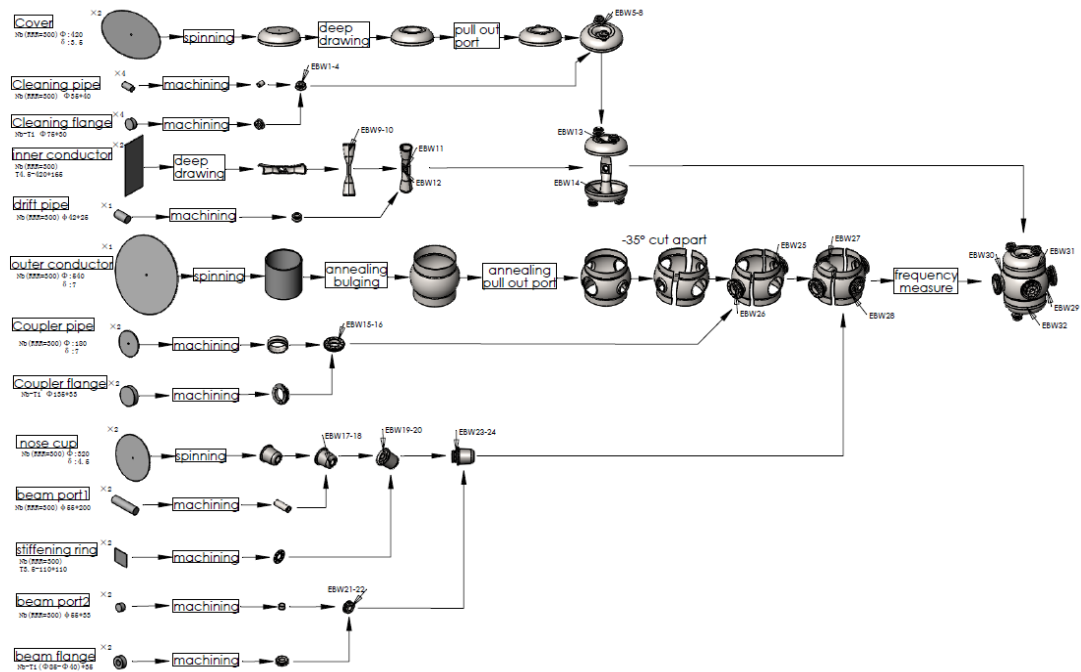

Fig. 11: The fabrication sequence of the HWR

## 4. Test

### 4.1 Post processing

The post processing of 325MHz HWR is including ultrasonic cleaning, BCP, annealing, high pressure rinsing (HPR), clean assembly and low temperature baking. The post processing sequence and the HRP are shown in Fig.13 and Fig.14.

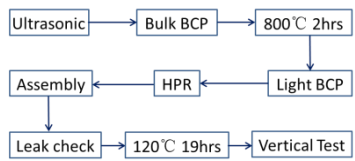

Fig. 13: The post processing of the HWR.

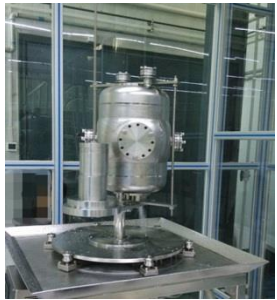

Fig. 14: The HPR of the HWR.

### 4.2 Vertical test

In the vertical test, the forward coupler and pick-up coupler are fixed length antenna, and the external Q can be selected to $1 \times 10^9$ and $3 \times 10^{10}$ respectively. The test results at RT and 4.2K are shown in Fig.15.

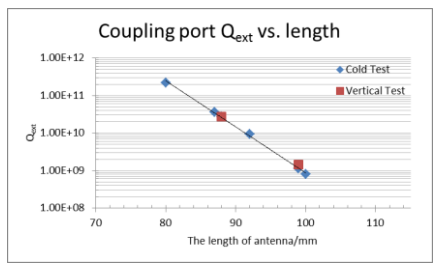

Fig. 15: The $Q_{ext}$ versus antenna length of the HWR.

Temperature sensors was connected to the up, center and down of the cavity to detect temperature changes caused by insufficient cooling. LHe level sensors, helium gas pressure sensors, cavity vacuum gauges and X-ray radiation was also online monitored. A 1 kW solid-state amplifier and LLRF control system is necessary to the vertical test. The HWR cooled down in dewar is shown in Fig.16.

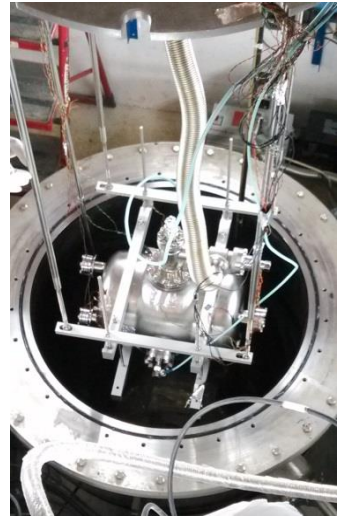

Fig. 16: The HWR and dewar.

In vertical test, the multipacting effect was occurred in low field（0～5MV/m） and high field (10.5MV/m, 13～15 MV/m) at 4.2K. The multipacting phenomenon is shown in Fig.17. RF conditioning can overcome the multipacting and improve cavity performance. After an hour of RF conditioning, the multipacting barriers were soft and reduced greatly. At 2K, after several hours of RF conditioning, the multipacting was insurmountable, so the test had to be stopped. The reason still needs to be further studied.

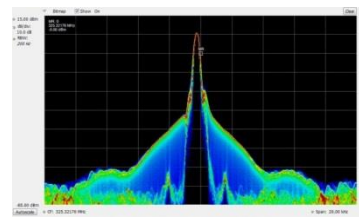

Fig. 17: The multipacting spectrum during VT aging.

At 4.2K, 325MHz HWR performance reached $Q_0 = 1.4 \times 10^9$ at $E_{acc} = 7 MV/m$ and $Q_0 = 4.3 \times 10^8$ at $E_{acc} = 15.9 MV/m$. The maximum peak fields reached 66.2MV/m and 77.6mT, which limited by field emission. The X-ray appeared at 11 MV/m, and the test result of the $Q_0$ vs. $E_{acc}$ is shown in Fig. 18.

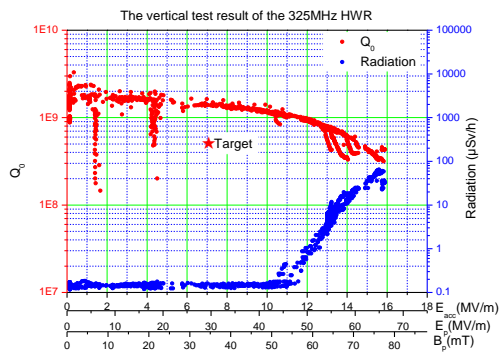

Fig. 18: The vertical test result of the HWR.

## 5. Summary

A 325MHz β=0.14 HWR prototype has been successfully developed for CW high beam proton accelerator. The EM parameters of the HWR have optimized a good result: $E_{peak}/E_{acc}$=4.2, $B_{peak}/E_{acc}$=4.9mT/(MV/m), R/Q=195Ω and G=74Ω, at the expense of reducing effective longitudinal space, which should be find a trade-offs between them for designer in engineering applications. The optimized mechanical design makes the cavity obtain a reasonable tuning range, low $df/dp$ and LFD. The cavity performance reached $E_{acc} = 15.9 MV/m$ with $Q_0 = 4.3 \times 10^8$ at 4.2K, and the curve of $Q_0$ vs. $E_{acc}$ is fairly flat. For the maximum peak fields ($E_{peak}$, $B_{peak}$) result is not too high, there are many probable improvement by future processing. All the multipacting barriers during VT are soft, and consistent with the simulation result. In next scheme, further surface processing (including roll grinding and polishing, plasma cleaning) will be done for better HWR's performance, and will be tested at 2K.